\documentclass{article}
\title{Pessimistic Testing}
\author{Ernie Cohen\\ Microsoft}

\usepackage{ifthen}
\usepackage{latexsym}
\usepackage{amssymb}
\usepackage[margin=4cm]{geometry}
\def\/{\vee}

\def\Def#1{{\em #1}}

\def\Prestate#1#2#3{\mathrm{prestate}(P,o,s)}

\def\Log#1{\mathrm{log}(#1)}

\begin{document}
\maketitle
\begin{abstract}
We propose a new approach to testing conformance to a nondeterministic
specification, in which testing proceeds only as long as increased test coverage
is guaranteed.
\end{abstract}

In testing that a system meets a nondeterministic specification
\cite{OptimalTesting}, it is usually assumed that the system is fair
to each transition of the specification (i.e., the system will make
every possible nondeterministic choice if given enough opportunities).
But in fact, some transitions might be unlikely or impossible for a given
implementation.  When this happens, common model-based testing
practices (e.g., following a precomputed tour of the
state space) often lead to wasted test cycles and poor test coverage.
We propose an alternative approach, in which the tester uses a
dynamically computed strategy that is guaranteed to eventually
increase coverage; testing stops as soon as the system has a strategy
to avoid further coverage.

As usual, we cast the test problem as a game; here, it is conveniently
represented as a (directed) hypergraph. A \Def{hypergraph} is given by
a set of vertices and a set of (hyper)edges. Each edge is given by a
head vertex and a set of tail vertices; we say it is \Def{incident} to
its head.  An edge is \Def{reachable} iff all of its tail vertices are
reachable, and a vertex is reachable iff one of its incident edges is
reachable (as usual, taking the minimal solution).  The \Def{rank} of
a reachable edge is the maximum of the ranks of its tail vertices (0
if the tail is empty), and the rank of a reachable vertex is one plus
the minimum rank of its reachable incident edges.

In the test context, the hypergraph vertices are system states, each
hyperedge represents a possible test stimulus, the head of the
hyperedge is the state in which the stimulus can be delivered, and the
tail of the hyperedge gives the states to which the system is allowed
to transition under the stimulus.  To keep track of which states have
been explored by the test, we add trivial edges (with empty tails)
incident on each state (other than the initial state).  When the
system first visits a state, this incident edge is removed,
``marking'' the state.  Thus, the test state consists of a hypergraph
and a current state (an unmarked vertex of the hypergraph), and a move
of the testing game consists of the tester choosing an edge incident
on the current state and the system choosing a new current state from
the tail of the edge (marking the state if it is unmarked).  Coverage
is measured as the number of marked states.

A key observation is that the tester has a strategy to increase
coverage iff the current state is reachable.  If the current state is
unreachable, the system can prevent further marking by always
choosing an unreachable successor state.  Conversely, if the current
state is reachable, the tester's strategy is to always choose a
reachable edge incident on the current state of minimal rank; this
results in either movement of the system to a state of lower rank or
marking of the new state. Since rank is bounded below by 0, some
state is eventually marked. (In fact, this strategy is optimal in
the sense of minimizing the upper bound on the number of moves before
the next marking.)

To operationalize this test strategy, we need to maintain (under game
moves) the ranks of reachable edges incident on the current state.
The obvious decremental hypergraph reachability algorithm
\cite{ShortestPath} maintains ranks for all states and edges in total
time $O(E + S\cdot H)$, where $E$ is the number of states marked, $S$
is the total number of states, and $H$ is the size of the hypergraph.
However, since ranks can only increase, and this algorithm updates
ranks in increasing rank order, we can delay updating ranks for edges
and states whenever they exceed that of the current system state.  We
can also delay adding to the graph edges incident on unmarked states
(``dead'' edges).  This improves the worst-case complexity to $O(E +
R\cdot H')$, where $R$ is the maximum rank assumed by the system state
prior to termination (i.e., the maximum number of stimuli between
explorations), and $H'$ is the total size of the live edges.  In the
worst case $R$ is $E$, but in practice, $R$ grows much more slowly in
$E$; for example, for random graphs of bounded degree, $R$ is
$O(\Log{E})$.  In experiments testing conformance of the Microsoft
Hypervisor to its functional specification \cite{HyperV}, the
hypergraph computation is dominated by test instrumentation, even
on tests with over $10^5$ states.

There are several potentially useful transformations of the test
hypergraph.  First, other kinds of test coverage (edge coverage,
branch coverage, etc.) can be obtained using standard techniques.
Second, in the common case where an operation can fail without a state
change, the head of the hyperedge appears in the tail, which makes the
edge unusable in the strategy above. We typically want to allow such
an edge to be used as if the failure were impossible, at least until a
test hits the failing case at least once; we can achieve this by
adding a new state and edge to break the self-loop.  Third, since the
worst-case computational cost grows with the maximum rank assumed by
the system state, it can be advantageous to compress a long 
sequence of preparatory operations into a single edge.

Finally, because dead edges are not added to the graph during
computation, the method here is fully compatible with lazy generation
of the state space, where edges incident on a state are generated
only when the state is marked.


\begin{thebibliography}{9}
\bibitem{OptimalTesting}[1]
Lev Nachmanson, Margus Veanes, Wolfram Schulte, Nikolai Tillmann,
Wolfgang Grieskamp.
\textsl{Optimal strategies for testing nondeterministic systems}. ISSTA 2004: 55-64

\bibitem{ShortestPath}[2]
G. Ausiello, P. G. Franciosa, and D. Frigioni,
\textsl{Partially Dynamic Maintenance of Minimum Weight Hyperpaths.}
J. of Discrete Algorithms, 3(1):27-46, 2005.

\bibitem{HyperV}[3]
\textsl{Microsoft Hypervisor Functional Specification.} Available from 
www.microsoft.com.

\end{thebibliography}
\end{document}